# Title: Signatures of moiré-trapped valley excitons in $MoSe_2/WSe_2$ heterobilayers

**Authors:** Kyle L. Seyler[*1], Pasqual Rivera[*1], Hongyi Yu[2], Nathan P. Wilson[1], Essance L. Ray[1], David Mandrus[3,4,5], Jiaqiang Yan[3,4], Wang Yao[#2], Xiaodong Xu[#1,6]

**Affiliations:**
[1]Department of Physics, University of Washington, Seattle, Washington 98195, USA.
[2]Department of Physics and Center of Theoretical and Computational Physics, University of Hong Kong, Hong Kong, China.
[3]Materials Science and Technology Division, Oak Ridge National Laboratory, Oak Ridge, Tennessee, 37831, USA.
[4]Department of Materials Science and Engineering, University of Tennessee, Knoxville, Tennessee, 37996, USA.
[5]Department of Physics and Astronomy, University of Tennessee, Knoxville, Tennessee, 37996, USA
[6]Department of Materials Science and Engineering, University of Washington, Seattle, Washington, 98195, USA.

*These authors contributed equally to the work.
[#]Correspondence to xuxd@uw.edu; wangyao@hku.hk

**Abstract: The creation of moiré patterns in crystalline solids is a powerful approach to manipulate their electronic properties, which are fundamentally influenced by periodic potential landscapes. In two-dimensional (2D) materials, a moiré pattern with a superlattice potential can form by vertically stacking two layered materials with a twist and/or finite lattice constant difference. This unique approach has led to emergent electronic phenomena, including the fractal quantum Hall effect[1–3], tunable Mott insulators[4,5], and unconventional superconductivity[6]. Furthermore, theory predicts intriguing effects on optical excitations by a moiré potential in 2D valley semiconductors[7–9], but these signatures have yet to be experimentally detected. Here, we report experimental evidence of interlayer valley excitons trapped in a moiré potential in $MoSe_2/WSe_2$ heterobilayers. At low temperatures, we observe photoluminescence near the *free* interlayer exciton energy but with over 100 times narrower linewidths (~100 μeV). The emitter *g*-factors are homogeneous across the same sample and only take two values, -15.9 and 6.7, in samples with twisting angles near 60° and 0°, respectively. The *g*-factors match those of the free interlayer exciton, which is determined by one of two possible valley pairing configurations. At a twist angle near 20°, the emitters become two orders of magnitude dimmer, but remarkably, they possess the same *g*-factor as the heterobilayer near 60°. This is consistent with the Umklapp recombination of interlayer excitons near the commensurate 21.8° twist angle[7]. The emitters exhibit strong circular polarization, which implies the preservation of three-fold rotation symmetry by the trapping potential. Together with the power and excitation energy dependence, all evidence unambiguously points to their origin as interlayer excitons trapped in a smooth moiré potential with inherited valley-contrasting physics. Our results open opportunities for 2D moiré optics with twist angle as a unique control knob.**

**Main Text:**

Heterobilayers formed by monolayer semiconducting transition metal dichalcogenides are a fascinating system for exploring nanoscale semiconductor optoelectronics with valley functionality[10,11]. Vertically stacked $MoSe_2$ and $WSe_2$ monolayers, for example, exhibit an atomically sharp interface and type-II band alignment which hosts interlayer excitons—the Coulomb bound state between electrons and holes located in different monolayers. Under circularly polarized optical pumping, electrons and holes preferentially populate the ±K valleys in the $MoSe_2$ conduction band and $WSe_2$ valence band, respectively. This allows for the formation of interlayer excitons with a plethora of appealing properties, such as valley-contrasting physics[7–10,12–14], long population and valley lifetimes[10,15–17], high electrical tunability[10,15,18], and strong many-body interactions[10,15,16].

One enticing possibility, which has so far only been theoretically explored, is to harness the spatially periodic moiré superlattice potential for excitonic manipulation. Recent scanning tunneling microscopy/spectroscopy measurements have directly measured the moiré pattern of 2D semiconductor heterobilayers, showing a periodically varying interlayer separation and electronic band gap[19]. The periodicity of this moiré superlattice is determined by the lattice constant mismatch and twist angle ($\theta$) between the layers (Fig. 1a). When the moiré period is larger than the interlayer exciton Bohr radius (~1 nm), the excitons will experience a spatially periodic potential modulation, forming a solid-state analogue of a bosonic quantum gas in an optical lattice[20]. The moiré potential minima can function as a smooth quantum-dot-like confinement potentials (Fig. 1b), and unique to these moiré-defined trapping sites is the preservation of three-fold rotational ($C_3$) symmetry[8,9]. Therefore, the moiré-trapped interlayer excitons should inherit valley-contrasting properties, a feature which distinguishes them from excitons bound to other randomly formed potential traps (e.g. defects, impurities, strain, etc.).

In this work, we report experimental signatures of interlayer excitons trapped in a moiré potential in $MoSe_2/WSe_2$ heterobilayers. The samples consist of exfoliated monolayers of $MoSe_2$ and $WSe_2$, stacked using a dry-transfer technique and aligned deterministically with $1^o$ uncertainty, which are encapsulated between ~10 to 30 nm hexagonal boron nitride to provide an atomically smooth substrate. Data from two devices are presented in the main text. Each device contains two heterobilayers that were simultaneously fabricated from the same pair of monolayers, but have regions with different $\theta$ (see Methods). This procedure minimizes the uncertainties when comparing the interlayer exciton properties at different twist angles. Device 1 contains heterobilayers with $\theta = 2^o$ and $20^o$ (Fig. 1c), while device 2 has regions with $\theta = 6^o$ and $57^o$.

$MoSe_2/WSe_2$ heterobilayers with aligned crystallographic axes generally exhibit bright interlayer exciton photoluminescence (PL) around 1.3 to 1.4 eV with a typical linewidth $\gtrsim$10 meV. Figure 1d shows PL spectra from the $\theta = 2°$ and 20° regions of device 1 under 5 µW 632.8 nm HeNe laser excitation (beam spot size ~ 1 $\mu m^2$) at a temperature of 1.6 K. In both heterobilayer regions, the intralayer exciton PL near 1.65 eV is strongly quenched relative to PL from the isolated monolayers due to ultrafast interlayer charge transfer[21]. However, there is a stark contrast in the interlayer exciton PL intensity, with the $2^o$ region being over two orders of magnitude brighter than the 20° region. This is due to the large mismatch of the first Brillouin zone corners between electrons and holes in 20° heterobilayer, which strongly suppresses the PL quantum yield compared to the nearly aligned 2° sample.

Under much lower excitation power (20 nW) near the monolayer $WSe_2$ A exciton resonance (1.72 eV), the broad interlayer PL develops into several narrow peaks near the free interlayer exciton energy around 1.33 eV (Fig. 1e and Extended Data Fig. 1), which fit well to Lorentzian curves (Fig. 1e, inset). The average linewidth of the observed peaks is about 100 μeV, which is comparable to the quantum emitters reported in monolayers of $WSe_2$ (ref. 22–25) and hexagonal boron nitride[26], and two orders of magnitude narrower than previous reports of interlayer exciton PL[10,12–17,27,28]. Narrow PL peaks and power broadening were also observed in the 57° and 20° sample (Extended Data Fig. 1).

The evolution of the broad interlayer PL peak into several narrow lines at low power suggests that the interlayer excitons are trapped in confinement potentials. The strong power saturation and broadening is characteristic of trapped excitons, where under high-power excitation, the traps are filled, and the emission becomes dominated by delocalized excitons that have broadened linewidth. Moreover, the narrow-line emission is suppressed above 30 K, after which the broader PL peaks dominate the spectrum (Extended Data Fig. 2). To substantiate the assignment of these features to interlayer excitons, we performed low-power PL excitation (PLE) spectroscopy, scanning a continuous-wave laser from 1.55 to 1.77 eV while monitoring the intensity of the narrow PL peaks. The PLE spectrum in Fig. 1f features two prominent resonances, consistent with the absorption of the $MoSe_2$ and $WSe_2$ intralayer excitons, which establishes the interlayer exciton character of the narrow PL lines.

The trapped interlayer excitons exhibit strong valley polarization. Figure 2 shows circular polarization-resolved PL spectra under $\sigma^+$ excitation for heterobilayers with $\theta = 57°$, 20°, and 2°. Denoting the valley polarization $\eta \ = \frac{I_+ - I_-}{I_+ - I_-}$ , where $I_\pm$ is the $\sigma^\pm$ PL intensity, the narrow peaks in the 57° and 20° heterobilayers display over 70% valley polarization (Figs. 2a and b). On the other hand, the selection rule is reversed for the 2° sample, with $\sigma^-$ emission dominating under $\sigma^+$ excitation (i.e. $\eta < 0$). Using $\sigma^-$ excitation, we confirmed the time-reversal process, establishing the observation of co-circularly polarized PL for 57° and 20° and cross-circularly polarized PL for 2° (Extended Data Fig. 3). Furthermore, under linearly polarized excitation, no significant linear PL polarization components were detected (Extended Data Fig. 4). Additional samples near 0° and 60° stacking show similar narrow PL features and polarization properties (Extended Data Fig. 3).

The circular polarization properties of the trapped interlayer excitons are distinct from the quantum emitters reported in monolayer materials, which exhibit either intrinsically linearly or un-polarized PL[22–25]. In those systems, high magnetic fields or trion formation under strong electron doping are required to restore modest circularly polarized PL[22–25]. Fine structure splitting observed in linearly polarized monolayer emitters implies anisotropy of the trapping potential, which breaks the three-fold rotational symmetry of the host lattice[22]. While the exact origin of the quantum emitters in monolayers is currently unclear, they are generally thought to arise from excitons bound to defects[29], impurities, or strain-induced potential traps[30,31]. However, unlike the monolayer case, the strong circularly polarized PL from the trapped interlayer excitons implies that the confining potential must possess three-fold rotational symmetry. One possible origin for the observed exciton trapping in the heterobilayer is the potential landscape resulting from the moiré superlattice, which naturally forms arrays of confinement centers with local atomic configurations that maintain $C_3$ symmetry[9]. This preserves valley optical selection rules and generally allows for cross-circularly polarized emission by interlayer excitons[8,9,14,27,32], which will be discussed later.

To support the moiré potential assignment, we performed magneto-PL spectroscopy to determine the Landé $g$-factor of trapped interlayer excitons. In Figs. 3a-c, we show circularly polarized PL under linearly polarized excitation as a function of perpendicular magnetic field $B$ for heterobilayers with $\theta$ = 57°, 20°, and 2°, respectively. The linearly polarized excitation equally pumps the degenerate ±K valleys of the monolayers, resulting in equal intensity and energy for $\sigma^+$ and $\sigma^-$ PL at 0 T. For nonzero magnetic fields, $\sigma^+$ and $\sigma^-$ PL split strongly in energy due to the valley Zeeman effect[33–36], but otherwise maintain the same peak structures, as illustrated by the spectra at 3 T (top row of Fig. 3). We visualize the full magnetic field dependence by plotting the total PL intensity versus the emission energy and $B$ (middle row of Fig. 3). For a given twist angle, we observe that for *all* PL peaks, the energies of $\sigma^+$ and $\sigma^-$ emission shift equally and oppositely to one another. Since several of the peaks are closely spaced in energy (~1 meV or less), the large energy shifts cause crossing points between the $\sigma^+$ and $\sigma^-$ emission of different PL peaks at high magnetic fields.

The bottom row of Fig. 3 shows the energy difference between the $\sigma^\pm$ PL ($\Delta = E_{\sigma^+} - E_{\sigma^-}$, where $E_{\sigma^\pm}$ is the peak energy of $\sigma^\pm$ polarized PL) of representative trapped interlayer excitons in each twisted heterobilayer. The extracted $g$-factor of 6.72 $\pm$ 0.02 for $\theta$ = 2° heterobilayer is nearly the same as the free interlayer exciton in near 0° samples (Extended Data Fig. 5). For $\theta$ = 57°, the effective $g$-factor is -15.89 $\pm$ 0.03, which is also very close to the $g$-factor of -15.1 reported for free interlayer excitons in 54° twisted heterobilayer[12]. We found that the $g$-factor is not only uniform between different trapped interlayer excitons in the same heterobilayer, but also nearly the same for different heterobilayers with similar twist angle (Extended Data Fig. 6).

The valley magnetic moment plays a central role in the distinct $g$-factors between nearly 0° and 60° heterobilayers. The Zeeman shift of carriers in the valley semiconductors has three contributions (Figs. 4a, b): spin ($\Delta_s = 2s_z\mu_B B$), atomic orbital ($\Delta_a = l_i\mu_B B$), and a valley contribution ($\Delta_v = \tau\alpha_i\mu_B B$) from the Berry phase effect[33–36]. Here, $\tau = \pm 1$ is valley index, $s_z = \pm 1/2$ is electron spin index, $\mu_B$ is Bohr magneton, $\alpha_i$ is valley $g$-factor for the conduction ($i$ = $c$) or valence band ($i$ = $v$), and $l_i$ is the magnetic quantum number for band $i$ ($l_c = 0$ and $l_v = 2\tau$). The Zeeman splitting of the interlayer exciton is then the difference between the Zeeman shifts of conduction band edge in $MoSe_2$ and valence band edge in $WSe_2$.

For interlayer excitons with spin-conserving optical transitions, spin does not contribute to the exciton Zeeman shift. Furthermore, the atomic orbital contribution to the Zeeman splitting, $-4\mu_B B$, is independent of the twist angle. The major difference between nearly 0° and 60° stacking therefore lies in the valley magnetic moment contribution because of their distinct conduction-valence valley pairing. The bright interlayer exciton configuration can be uniquely specified by the valley index pair ($\tau_c$, $\tau_v$), which is $(+,+)$ or $(-,-)$ for nearly 0° stacking (Fig. 4a), and $(+,-)$ or $(-,+)$ for nearly 60° stacking (Fig. 4b). Consequently, for 0° stacking, the excitonic Zeeman splitting is similar to monolayers and may be written as $\Delta_0 = -2(2 - \Delta\alpha)\mu_B B$, where the valley contribution $\Delta\alpha = \alpha_c - \alpha_v$. For the 60° stacking case, the valley contribution changes from a cancellation to a sum[12], which gives rise to much larger Zeeman shift $\Delta_{60} = -2(2 + \Sigma\alpha)\mu_B B$, where $\Sigma\alpha = \alpha_c + \alpha_v$ is the summation of valley $g$-factors, and thus larger effective interlayer $g$-factor than 0° stacked heterobilayer (Supplementary Discussion).

From above analysis, we see that interlayer exciton $g$-factor is a fingerprint of their valley configuration and valley magnetic moment. The defect-localized excitons in monolayers do not possess valley-contrasting properties[22–25] since the bulk crystal structure is not retained in the extent of the exciton wavefunction. Indeed, excitons bound to defects or strain-confined potentials observed in $WSe_2$ have distinct $g$-factors (a few times larger) compared to the free intralayer exciton counterparts, and often exhibit fine structure splitting and absence of circular polarization from the anisotropic quantum confinement[22–25]. Our observation that the trapped interlayer excitons have the same $g$-factors and twist angle dependence as the free ones demonstrates that the trapping potential must be smooth enough and three-fold rotational symmetric to allow the inheritance of the valley properties from the heterobilayer bulk. These findings strongly suggest that extrinsic factors (defects, impurities, strain) cannot be the origin of the trapping, whereas the moiré potential traps are the only known candidate.

Another remarkable finding which further supports the above picture is that the $g$-factor of interlayer excitons for $\theta$ = 20° is -15.78 $\pm$ 0.05 (bottom row, Fig. 3c), which is nearly the same as that of $\theta$ = 57° heterobilayer. This, at the first glance, is counterintuitive, as 20° is closer to 0° than to 60°. The mystery is solved by noticing that 21.8° is a commensurate stacking angle[7] that produces the shortest superlattice periodicity, with a supercell of size $\sqrt{7}a \times \sqrt{7}a$. Figure 4c shows the conduction (solid points) and valence band edges (open points) of heterobilayer in the extended Brillouin zone scheme, where red and blue denote +K and -K valley respectively. For a random twist angle, the valleys from $MoSe_2$ and $WSe_2$ are in general not aligned in momentum space, and thus interlayer excitons are momentum-indirect and optically dark. At 21.8° twisting angle, the conduction and valence band ±K points are misaligned in the first Brillouin zone but overlap on the second Brillouin zone, with the $(+,-)$ and $(-,+)$ valley pairings. As a result, the interlayer exciton at this commensurate stacking can directly couple to light for radiative recombination, with the momentum mismatch from the twisting compensated by the reciprocal lattice vectors from the two layers (i.e. Umklapp recombination)[7]. Since the valley pairings of 21.8° are the same as for 60° twisted heterolayers, they have the same $g$-factor. The optical dipole of the Umklapp recombination is expected to be very weak compared to that of the direct recombination at 0° and 60°. Indeed, our measurement reveals the PL intensity of $\theta$ = 2° is about 100 times stronger than that of $\theta$ = 20° (Extended Data Fig. 7).

Our heterobilayer of $\theta$ = 20° forms an interesting concatenated moiré pattern, as schematically shown in Extended Data Fig. 8. Close-up views of any local region resemble the commensurate moiré pattern of the 21.8° stacking, but the interlayer translation varies smoothly over a longer length scale with the periodicity $\mathrm{A} = \frac{a}{\sqrt{7}\delta\theta}$. Here, $\delta\theta$ is the small deviation of the actual twisting angle from the commensurate angle of 21.8°, and A $\approx$ 4 nm at $\theta$ = 20°. In this concatenated moiré supercell, we can also identify three local regions that retain the three-fold rotational symmetry, which must correspond to either the minima or maxima of the moiré superlattice potential for the interlayer excitons (Supplementary Discussion). As with the $\theta$ = 2° and $\theta$ = 57° moiré pattern, interlayer excitons in our 20° heterobilayer trapped in such potential minima can retain circularly polarized valley optical selection rules, consistent with the high degree of PL polarization observed. The experimental observations of narrow-line emission, circular selection rules, and distinct binary $g$-factors at a variety of twisting angles are compelling evidence of interlayer excitons trapped in moiré superlattice potential.

Finally, we remark on some subtle features in the observed moiré-trapped interlayer excitons. The polarization of PL is determined by local crystal symmetry of moiré potential. Theory shows two high-symmetry points, A and B, in the moiré superlattice[8,9]. The former has co-circular optical selection rules, while the latter has circular polarization reversal. Our experiment implies the moiré exciton is located at the B point of the moiré supercell for $\theta$ = 2°, and the A point for $\theta$ = 20° and $\theta$ = 57°. This also explains why the $g$-factor we measured was positive for $\theta$ = 2° but negative for $\theta$ = 20° and $\theta$ = 57°, because the Zeeman splitting is defined as the energy difference between $\sigma^{\pm}$ PL (Supplementary Discussion).

The moiré-trapped interlayer excitons we observe are nonuniform in the number of emitters, their relative intensities, and their energies. We often detect several narrow PL peaks with sub-meV energy separation within the laser excitation spot. The intensities and valley polarization can also vary significantly between the peaks (Extended Data Fig. 9). These observations imply that the moiré superlattice potential has some inhomogeneity, which is not surprising because imperfections are expected during fabrication of the moiré superlattice. In fact, fabricating a homogenous moiré pattern by mechanical transfer is an open challenge in the community. On the other hand, different emitters for similar stacking configurations have a common $g$-factor, which is expected in a moiré potential trap but *not* in other extrinsic traps. As we mentioned, the $g$-factor of the moiré exciton is a property of the heterobilayer bulk determined by the valley pairing only, which is the same for every local region of a given moiré pattern. Lastly, considering that the energy spacing of several PL lines is on the same order as the repulsive interaction between proximate interlayer excitons, there is a possibility that some of the narrow PL peaks may originate from the cascaded emission of multi-exciton states in a single trap or several neighboring traps. All these possibilities require future studies with improved sample quality and possibly new scanning probe techniques, such as near-field scanning microscopy with sub-10 nm spatial resolution. Nevertheless, our observation of moiré excitons provides a promising starting point to explore several intriguing theoretical proposals related to quantum photonics, such as entangled photon sources, giant spin-orbit coupling[8], topological excitons[8,37], and much more.

## Methods:

### Sample fabrication

The samples were fabricated by dry-transfer of monolayers obtained from the mechanical exfoliation from bulk crystals. The crystal orientation of the individual monolayers was determined by linear-polarization-resolved second-harmonic generation[10] prior to transfer. During the transfer process for device 1, a region of the $WSe_2$ monolayer tore off and twisted 18° relative to the original piece (see Fig. 1c). This yield heterobilayers with $\theta$ = 2° and 20°. Device 2 was fabricated from a large $WSe_2$ monolayer and an $MoSe_2$ piece with two overlapping monolayers (monolayer-bilayer-monolayer). The two $MoSe_2$ monolayer regions are formed from different layers of a 2H-bilayer, so they were oriented 180° relative to one another prior to transfer. Therefore, after transfer, device 2 possessed two heterobilayer regions, one with near 0° twist angle (6°) and another close to 60° (57°). The difference in twist angle between the two heterobilayer regions deviates from the expected 60° because one of the $MoSe_2$ monolayer regions rotated slightly during transfer. To verify the absolute stacking orientation, we used a phase-resolved second-harmonic generation technique, as described in our previous work[38]. This yielded a reference heterobilayer sample with 56° twist angle, on which we measured a $g$-factor near -15.9 (Extended Data Fig 6c). The two regions of device 2 were then measured to have a $g$-factor of 6.7 and -15.9, which confirmed their different stacking orientations as well as the general correspondence between samples with twist angle near 0° or 60° and their $g$-factor. The absolute twist angle for samples with near 0° or 60° was ultimately determined by the $g$-factor (close to 6.7 for $\theta \approx 0°$ and close to -15.9 for $\theta \approx 60°$).

**Photoluminescence measurements**

PL measurements were performed in a home-built confocal microscope in reflection geometry, with the sample mounted in an exchange-gas cooled cryostat equipped with a 9 T superconducting magnet in Faraday configuration. The sample temperature was kept at 1.6 K, unless otherwise specified. A power-stabilized and frequency-tunable narrow-band continuous-wave Ti:sapphire laser ($M^2$ SolsTiS) was used to excite the sample, unless otherwise specified. The PL was spectrally filtered from the laser using a long-pass filter before being directed into a spectrometer, where the PL was dispersed by a diffraction grating (600 or 1200 grooves/mm) and detected on a Si charge-coupled-device. At the interlayer PL energies of ~1.4 eV, the spectral resolution was ~160 μeV for the 600 grooves/mm grating (Figs. 1d-e, 2b, 3b) and ~80 μeV for 1200 grooves/mm (Figs. 1e inset, 2a, 2c, 3a, 3c). Polarization-resolved PL data were acquired using a combination of quarter-wave plates, half-wave plates, and linear polarizers for excitation and collection.

**Acknowledgments:** We acknowledge J. Fonseca for fabrication assistance and Edo Waks for $g^{(2)}$ measurements. This work was mainly supported by the Department of Energy, Basic Energy Sciences, Materials Sciences and Engineering Division (DE-SC0018171). WY and HY were supported by the Croucher Foundation (Croucher Innovation Award), the RGC of Hong Kong (HKU17305914P), and the HKU ORA. DM and JY were supported by the US Department of Energy, Office of Science, Basic Energy Sciences, Materials Sciences and Engineering Division. XX acknowledges the support from the State of Washington funded Clean Energy Institute and from the Boeing Distinguished Professorship in Physics.

**Author Contributions:** XX, WY, KLS, and PR conceived the experiment. PR fabricated the devices, assisted by KLS and ELR. KLS and PR performed the measurements, assisted by NPW. KLS, PR, XX, WY, and HY analyzed and interpreted the results. JY and DGM synthesized and characterize the bulk crystals. KLS, XX, and WY, PR, and HY wrote the paper with inputs from all authors. All authors discussed the results.

**Competing Financial Interests:** The authors declare no competing financial interests.

**Data Availability:** The data that support the findings of this study are available from the corresponding author upon reasonable request.

## Figures

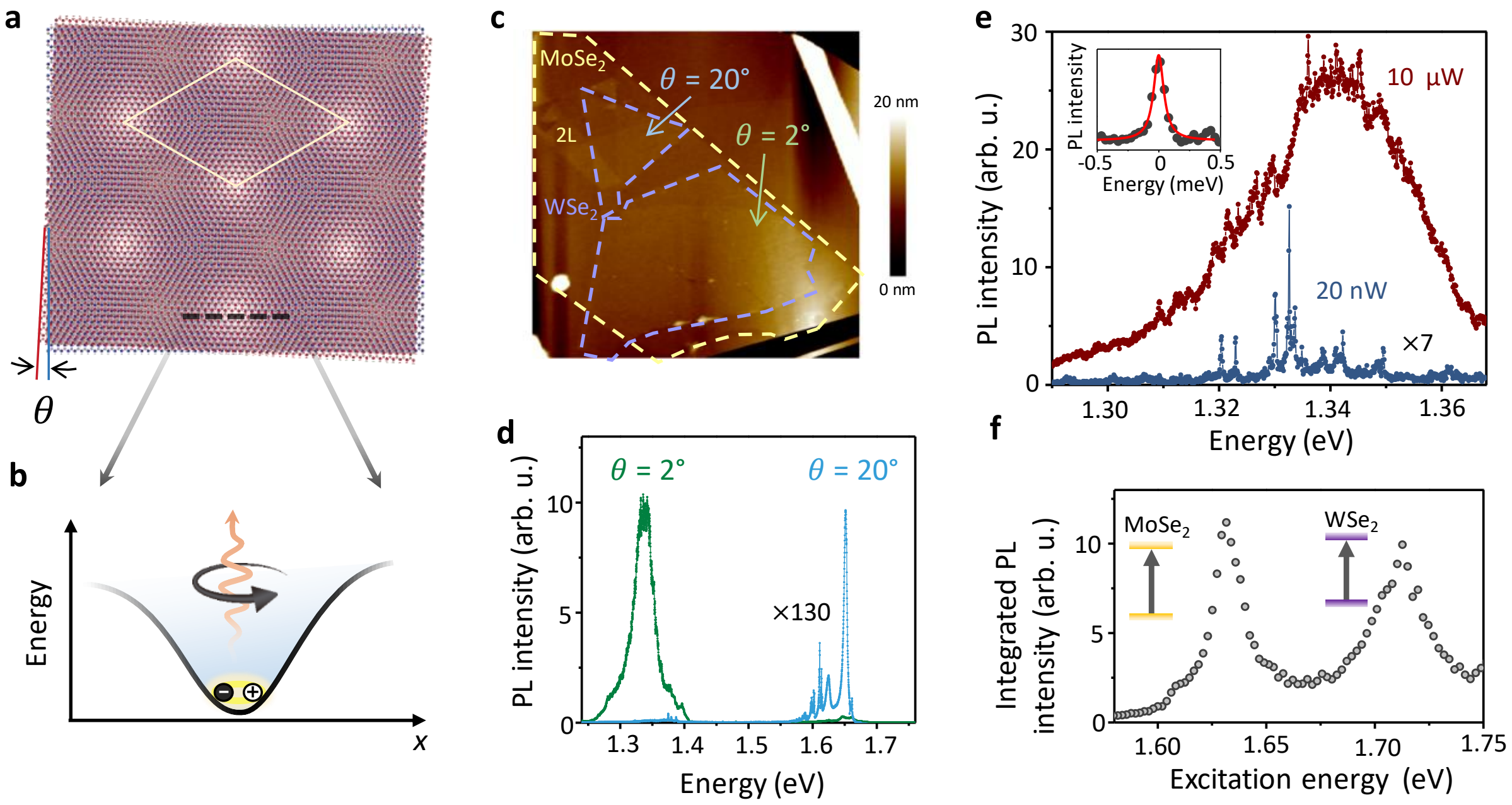

**Figure 1 | Moiré superlattice potential and observation of trapped interlayer excitons. a**, Illustration of the moiré superlattice formed in a heterobilayer with twist angle $\theta$. **b**, Cartoon of an exciton trapped in a moiré potential. **c,** Topographical height map of $MoSe_2$/$WSe_2$ heterobilayers encapsulated by hexagonal boron nitride, obtained from an atomic force microscope. Orange and purple solid lines indicate the $MoSe_2$ and $WSe_2$ monolayers, respectively. The heterobilayers have different twist angles, as indicated. **d**, Photoluminescence (PL) spectra from the heterobilayer with 2° (green) and 20° (blue, intensity scaled by 130×) twist angle, at 5 μW excitation power. **e**, Comparison of interlayer exciton PL from the 20° twist angle heterobilayer at 10 μW (orange) and 20 nW (purple, intensity scaled by 7×) excitation power. Inset, Lorentzian fit to a representative PL peak indicates a linewidth of ∼100 μeV (20 nW excitation power). **f**, PL excitation intensity plot on a narrow PL peak showing two prominent resonances corresponding to the intralayer exciton states in the monolayers of $MoSe_2$ and $WSe_2$. The intensity is integrated over a single peak of the $\theta$ = 57° sample, but the results are qualitatively similar for all other emission lines on each sample.

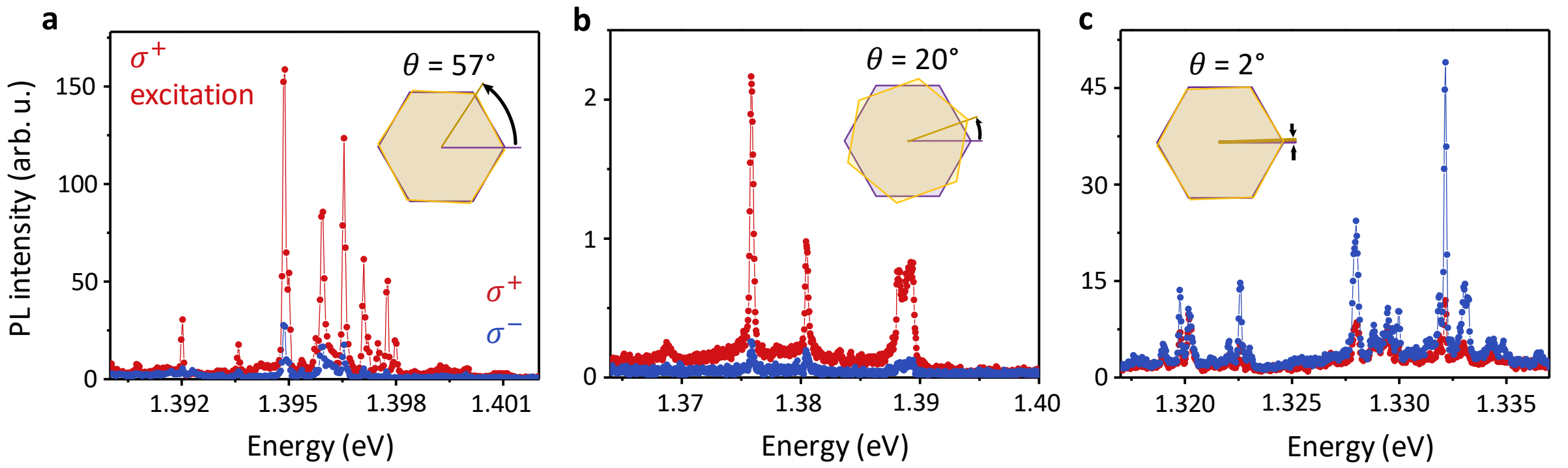

**Figure 2 | Valley polarization of trapped interlayer excitons.** Helicity-resolved photoluminescence (PL) spectra of trapped interlayer excitons of $MoSe_2/WSe_2$ heterobilayers with twist angle of **a**, 57°, **b**, 20°, **c** and 2°. Insets illustrate the twist angle of the three samples. The samples are excited with $\sigma^+$-polarized light at 1.72 eV. The $\sigma^+$ and $\sigma^-$ components of the PL are shown in red and blue, respectively. The PL from heterobilayers with twist angles of 57° and 20° are co-circularly polarized, while PL from the heterobilayer with twist angle of 2° is cross-circularly polarized.

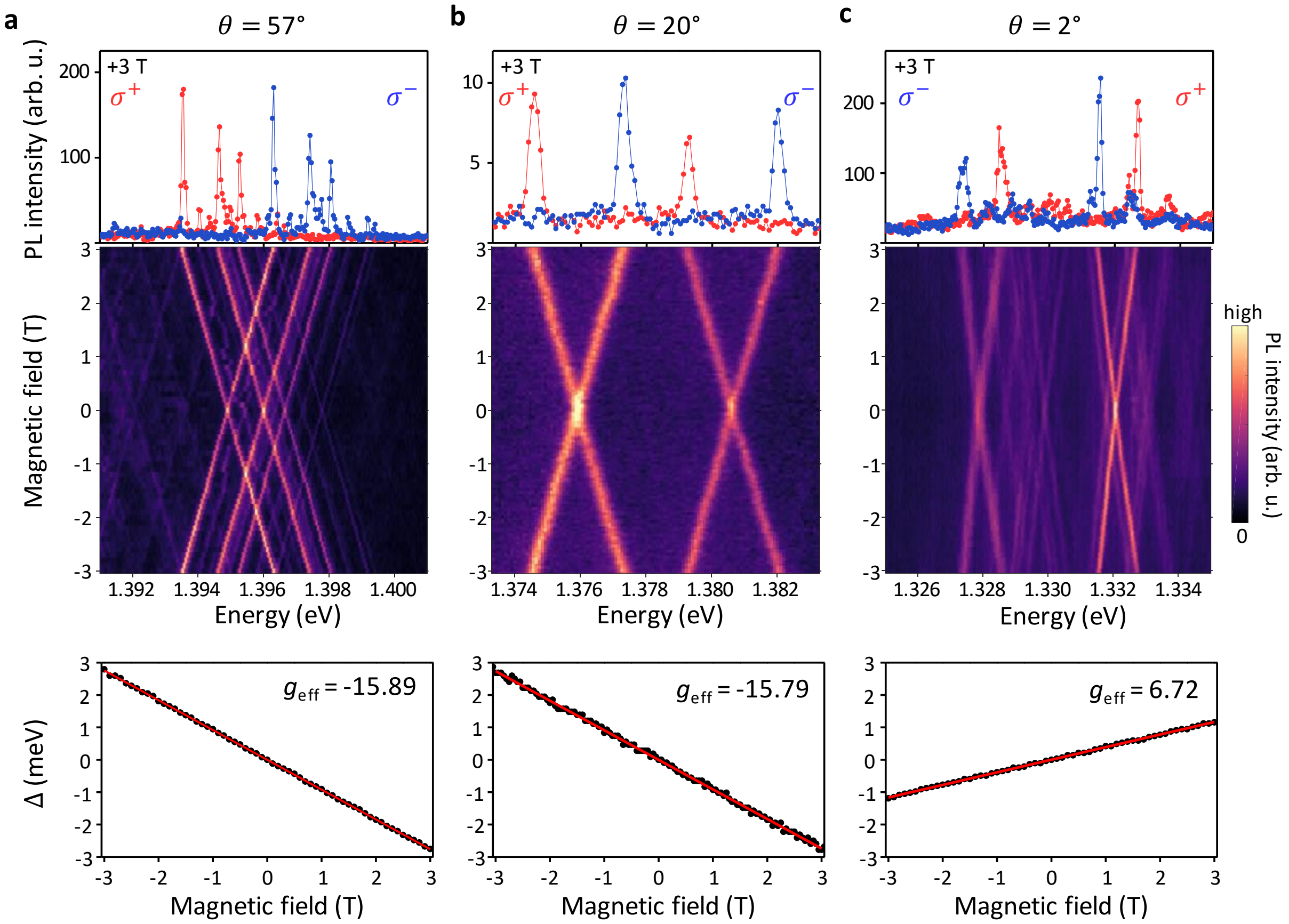

**Figure 3 | Twist-angle-dependent Zeeman splitting of trapped interlayer excitons.** Magnetic-field-dependent photoluminescence (PL) from interlayer excitons in the $MoSe_2/WSe_2$ heterobilayer with twist angle of **a**, 57°, **b**, 20°, **c** and 2°. Top row: Helicity-resolved PL spectra at 3T. The excitation is linearly polarized, and the $\sigma^+$ and $\sigma^-$ components of the PL are shown in red and blue, respectively. Middle row: Total PL intensity plot as a function of magnetic field, showing a linear Zeeman shift of the $\sigma^+$ and $\sigma^-$ polarized PL. Bottom row: Zeeman splitting of the polarization-resolved PL ($\Delta = E_{\sigma^+} - E_{\sigma^-}$) as a function of the applied magnetic field. The effective g-factors for the three samples ($-15.89 \pm 0.02$, $-15.79 \pm 0.05$, and $6.72 \pm 0.02$ for (**a**), (**b**), and (**c**), respectively) are extracted from a linear fit of $\Delta$ versus $B$ (red solid line).

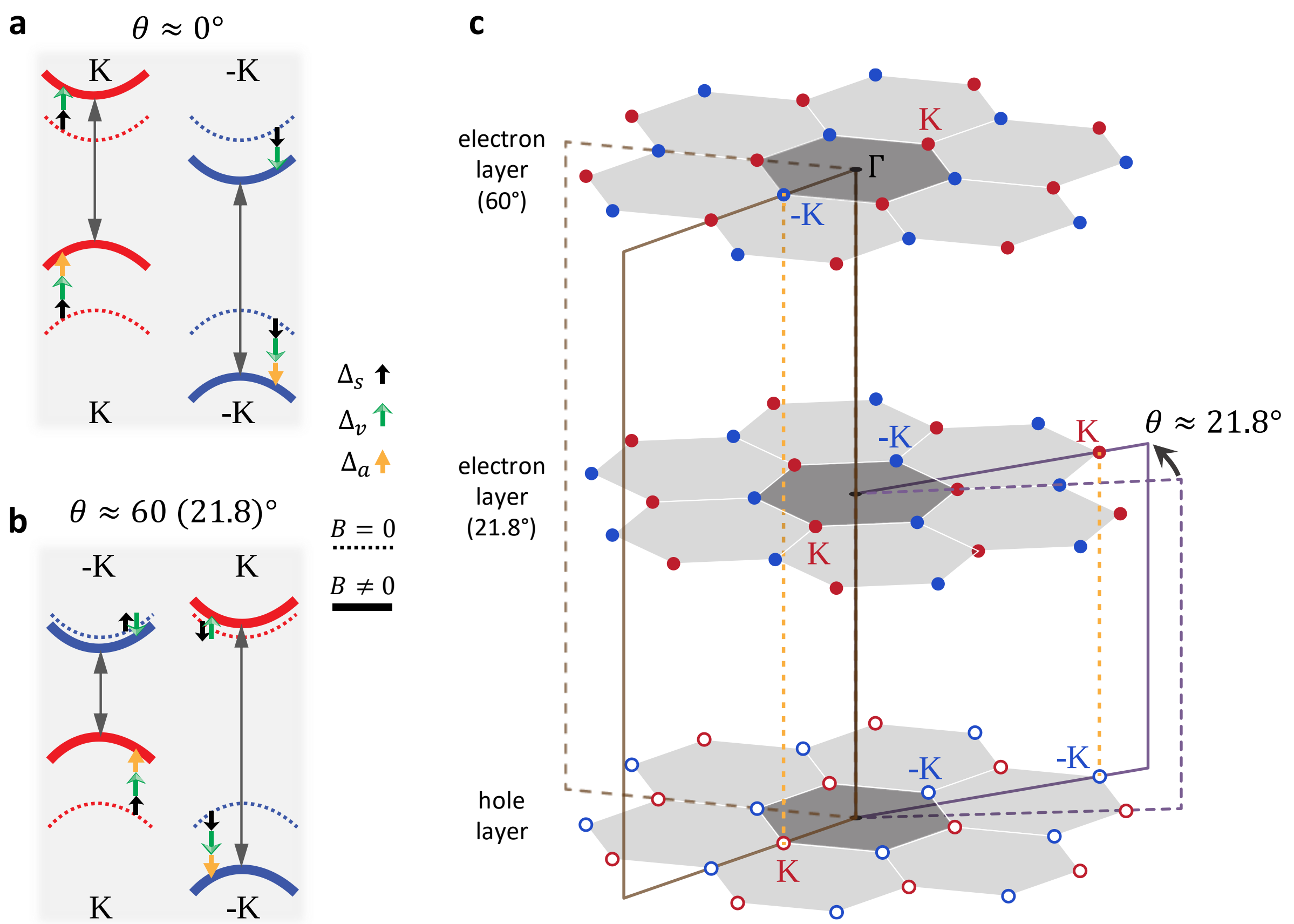

**Figure 4 | Twist-angle-dependent *g*-factors and Umklapp light coupling of interlayer excitons. a**, **b**, Energy level diagram showing the contributions of the interlayer exciton Zeeman shift by the electron spin ($\Delta_s$, in black), valley ($\Delta_v$, in green), and atomic orbital ($\Delta_a$, in orange) for the $MoSe_2/WSe_2$ heterobilayer with twist angle near 0° (**a**) and 60° (**b**). **c**, Schematic of valley alignments in extended Brillouin zone for a twisted heterobilayer. The open and solid points represent the $+$K (red) and $-$K (blue) valleys in the hole and electron layers, respectively. Near 21.8° twist angle, the $\pm$K and $\mp$K valleys align in the second Brillouin zone, which has the same valley pairing as the 60° twisted heterolayer and thus the same *g*-factor. Light coupling at 21.8° is facilitated by an Umklapp-type process.

# Extended Data Figures for

## Signatures of moiré-trapped valley excitons in $MoSe_2$/$WSe_2$ heterobilayers

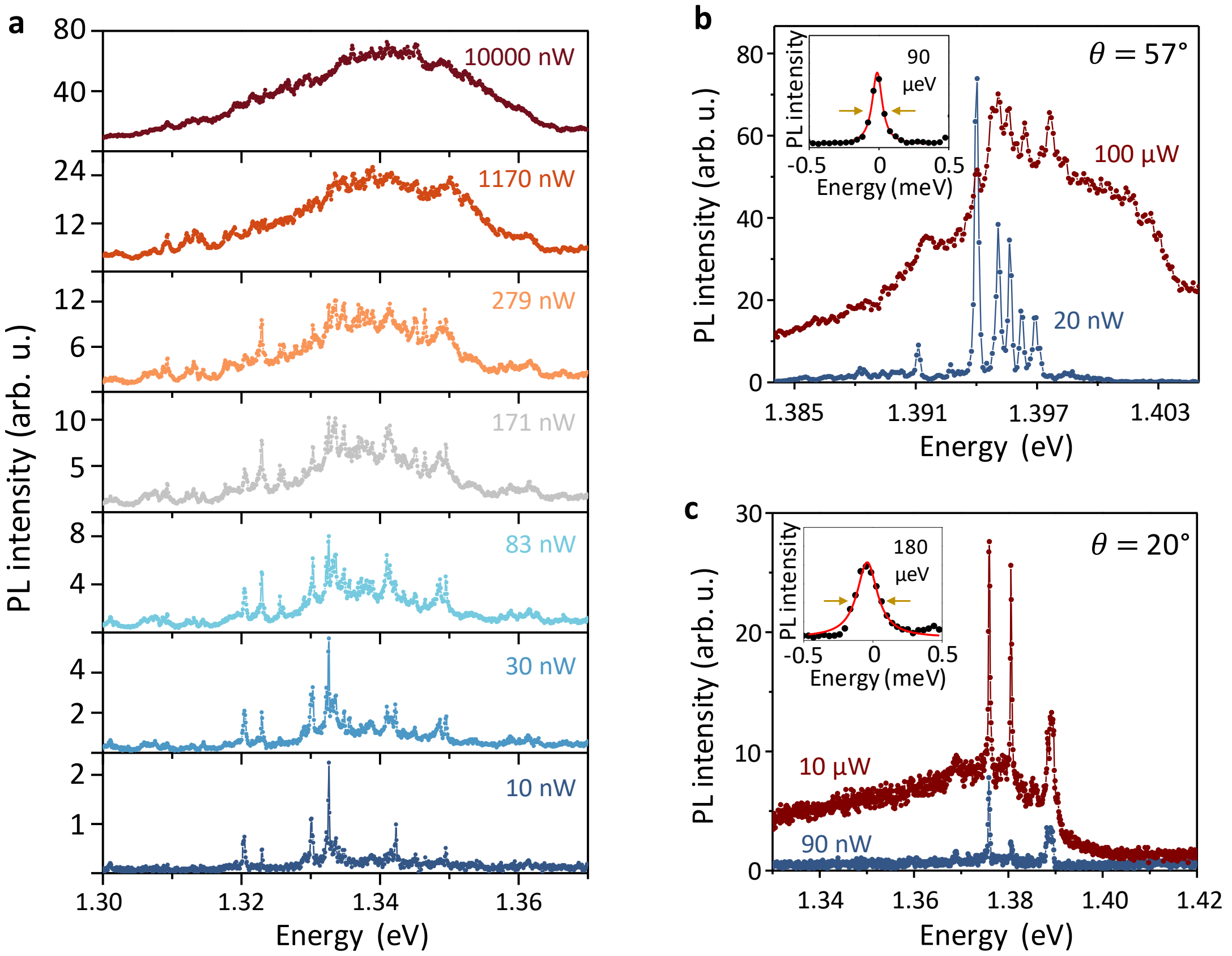

**Extended Data Fig. 1 | Supplementary power dependence data. a**, Interlayer exciton PL spectra at selected excitation powers for $\theta$ = 2° sample of device 1. **b**, **c**, PL spectra at low versus high power (indicated on the figure) for $\theta$ = 57° of device 2 (**b**) and $\theta$ = 20° of device 1 (**c**). Insets: Lorentzian fit to representative PL peaks with the indicated linewidths.

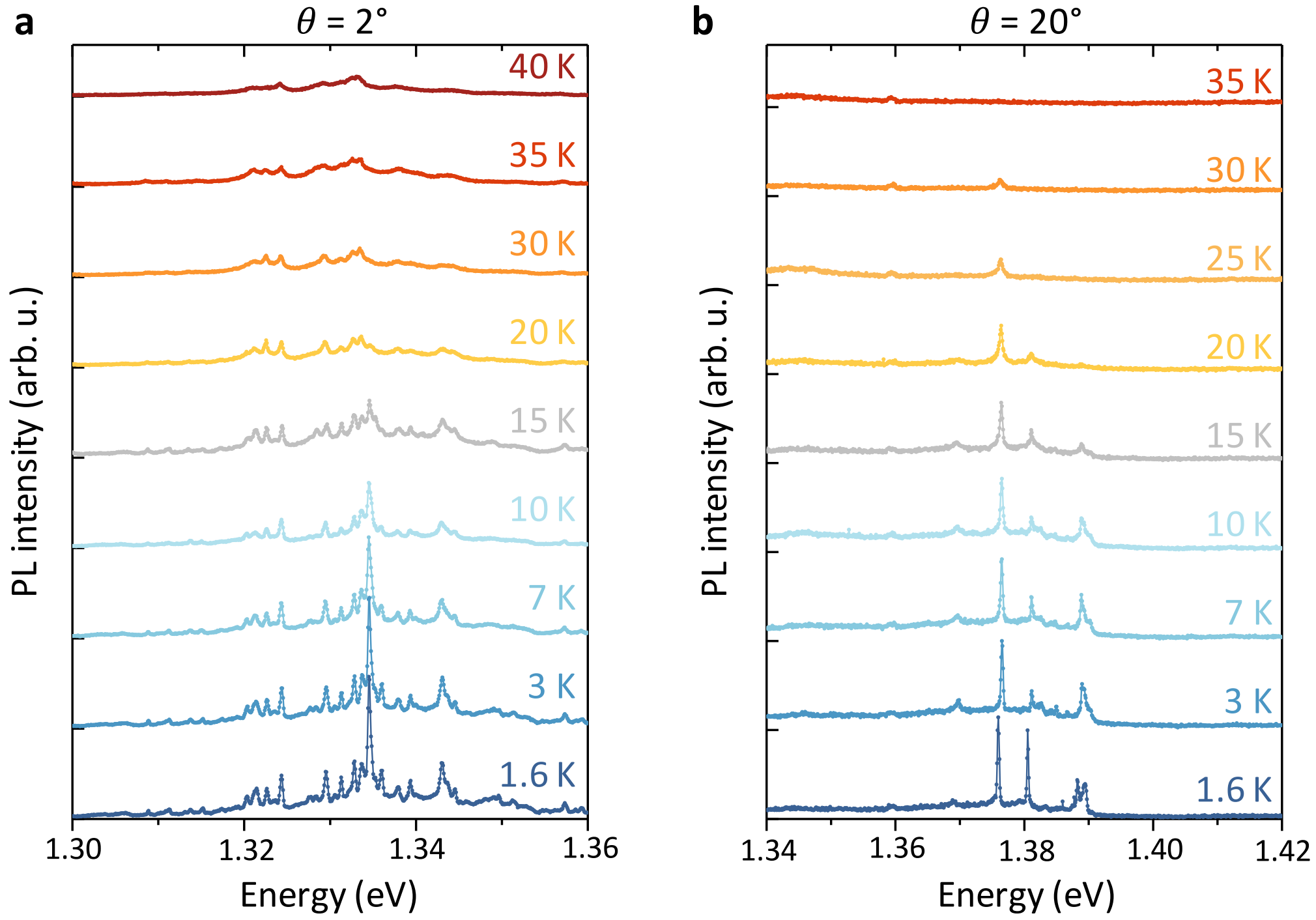

**Extended Data Fig. 2 | Temperature-dependent PL. a**, **b**, Temperature dependence of the interlayer exciton PL spectra for $\theta$ = 2° (**a**) and $\theta$ =20° (**b**) samples of device 1. The excitation powers were 20 nW (**a**) and 1 $\mu$W (**b**).

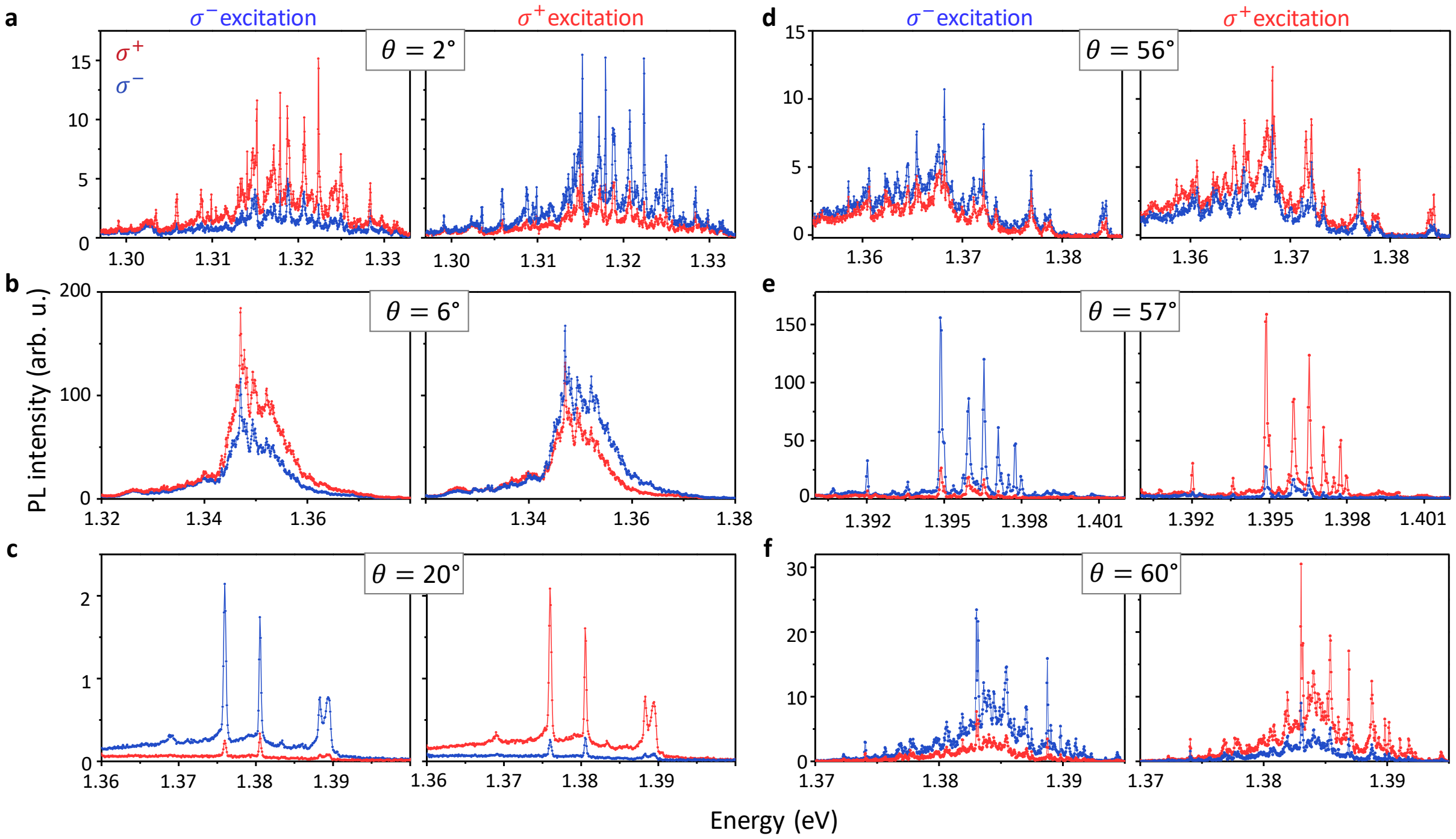

**Extended Data Fig. 3 | Supplementary circular-polarization-resolved PL spectra.** Each subfigure represents data from a different sample with the indicated twist angle. The left and right columns of each subfigure correspond to $\sigma^-$ and $\sigma^+$ polarized excitation. The red and blue curves indicate $\sigma^+$ and $\sigma^-$ polarized PL components. Spectra from the $\theta$ = 2° and $\theta$ = 20° region of device 1 in the main text are shown in (**a**) and (**c**), while $\theta$ = 6° and $\theta$ = 57° from device 2 are shown in (**b**) and (**e**). The $\theta$ = 2° spectra in (**a**) were acquired on a different sample region from that of the spectra in main text, which shows similarly strong polarization reversal. The spectra in (**d**) and (**f**) are from additional heterobilayers. The twist angle uncertainty is 1°.

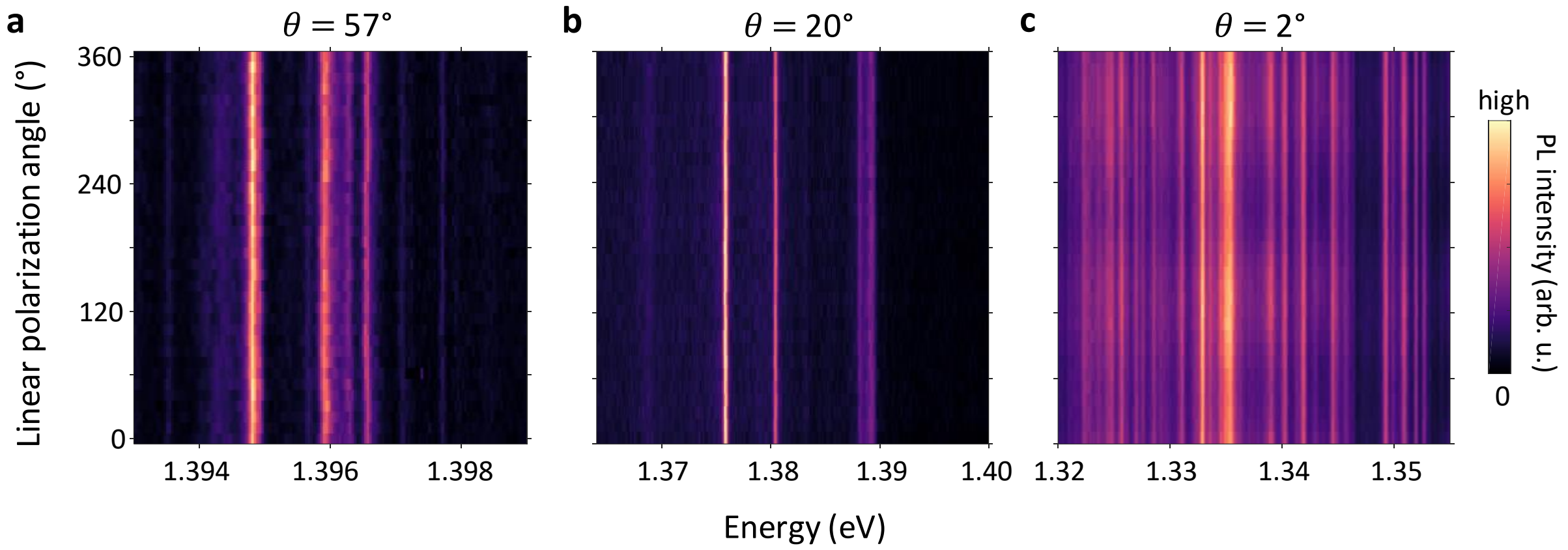

**Extended Data Fig. 4 | Linear-polarization-resolved PL. a**-**c**, PL intensity plots as a function of the linear polarization detection angle and photon energy under linearly polarized excitation for $\theta$ = 57° (**a**), 20° (**b**), and 2° (**c**). No linear PL polarization is observed.

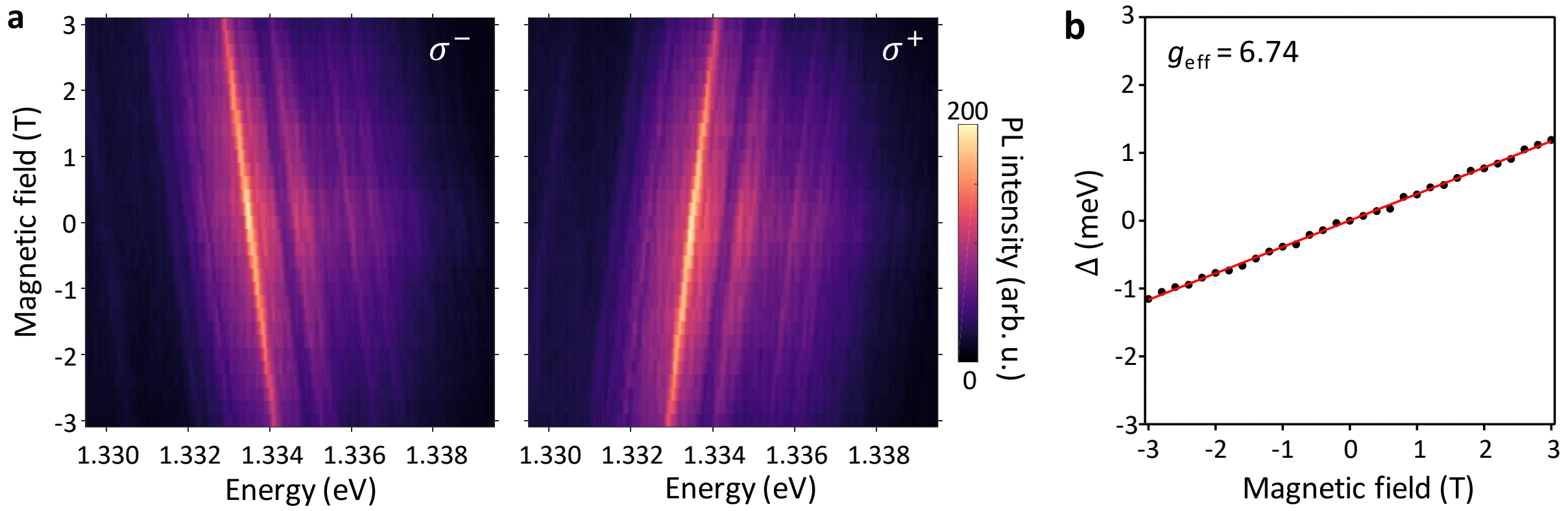

**Extended Data Fig. 5 | Free and trapped interlayer exciton *g*-factor for $\theta \sim 0°$.** The data is acquired on the $\theta$ = 6° region of device 2. **a**, PL intensity plot of $\sigma^+$ (right) and $\sigma^-$ (left) components as a function of applied magnetic field and photon energy. The broad background (free exciton) clearly shifts with same slope as the sharp trapped interlayer exciton on top. **b**, Valley splitting versus applied magnetic field, from which a *g*-factor of 6.74 $\pm$ 0.05 is extracted from a linear fit to Δ versus *B* (red solid line).

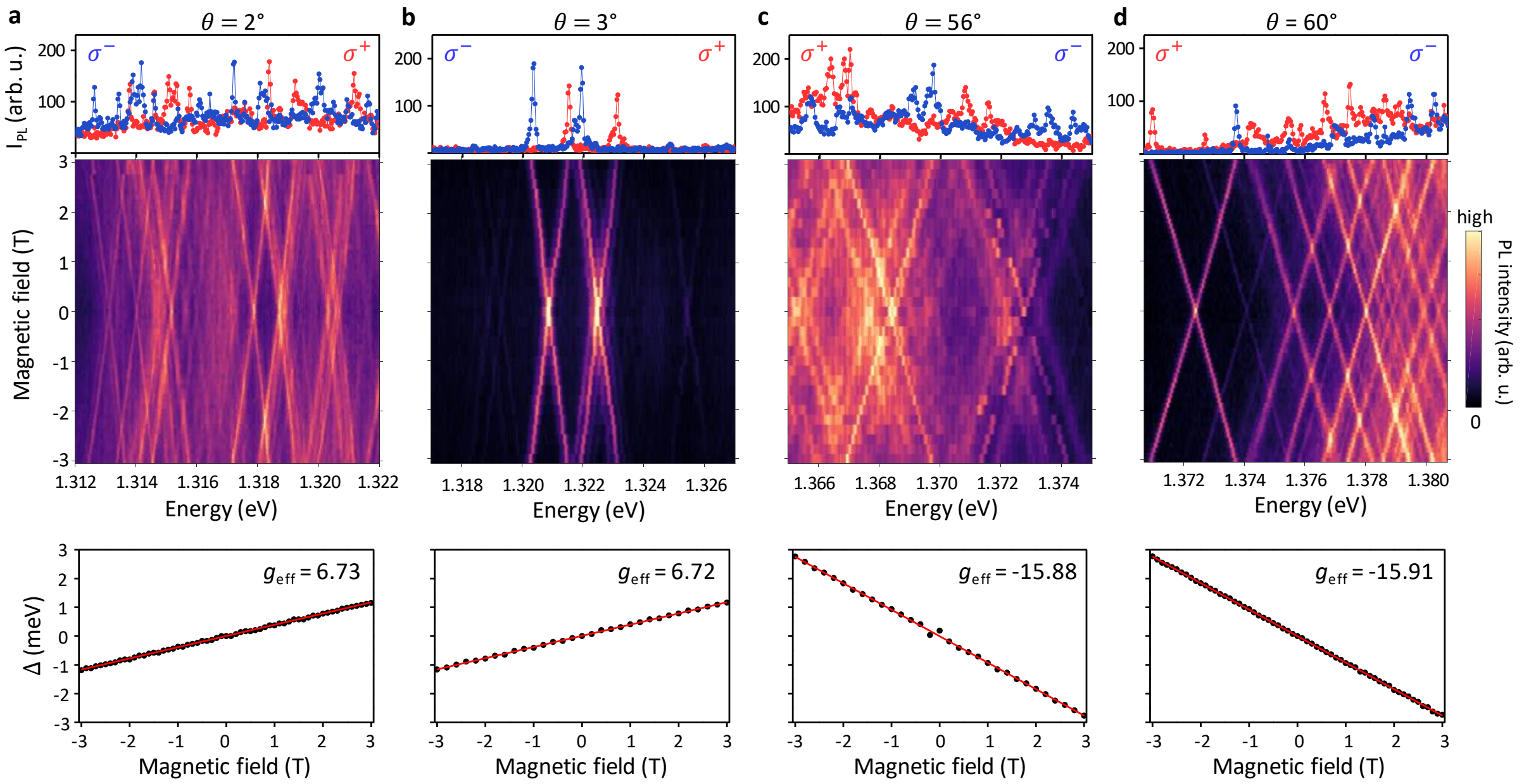

**Extended Data Fig. 6 | Uniformity of *g*-factors. a-d,** Magneto-PL of 2° (**a**), 3° (**b**), 56° (**c**), and 60° (**d**) heterobilayers. The data in the 2° heterobilayer (**a**) was taken at a different spot from the one in the main text. The excitation is linearly polarized. Top row: Circularly polarized PL spectra at 3 T. Middle row: PL intensity as a function of applied magnetic field, which display a clear linear Zeeman shift of the $\sigma^+$ and $\sigma^-$ components. Bottom row: Valley Zeeman splitting versus applied magnetic field. Linear fits (red solid lines) yield effective *g*-factors of heterobilayers at different twist angle, which are 6.73 ± 0.02, 6.72 ± 0.03, -15.88 ± 0.09, and -15.91± 0.02 for (**a**)-(**d**). The data confirm the *g*-factor uniformity across the same sample as well as different samples with similar twist angle.

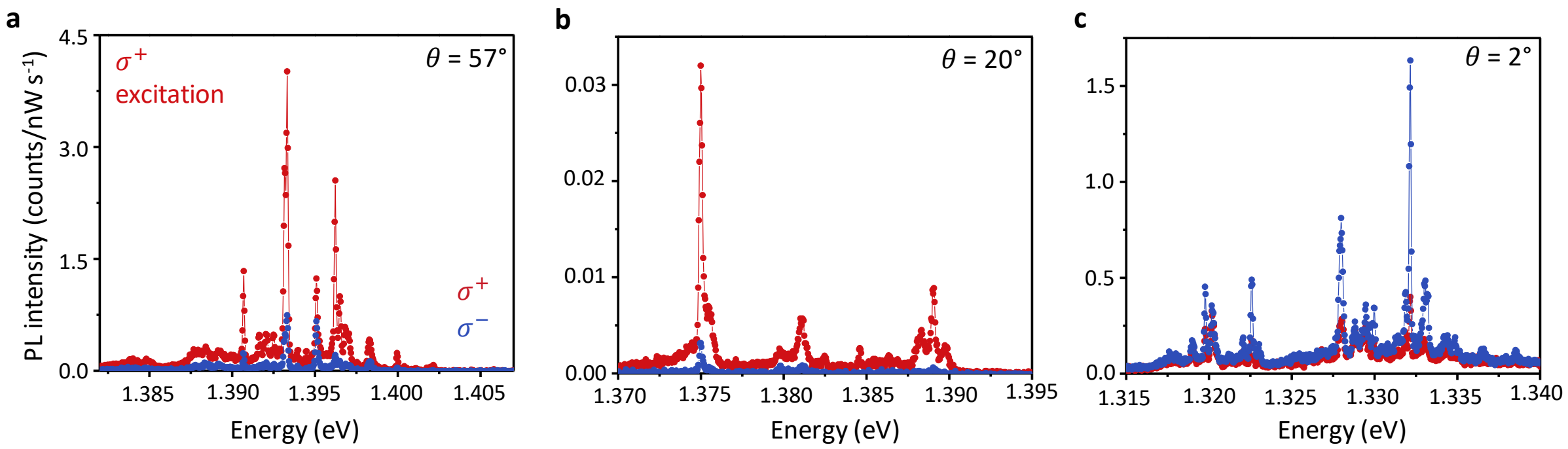

**Extended Data Fig. 7 | Calibrated PL spectra to compare PL intensity from heterobilayers with different twist angles.** The measurement reveals the PL intensity of $\theta = 2°$ and 57° is about 100 times stronger than that of $\theta = 20°$. Excitation powers were 10 nW for (**a**) and (**c**) and 100 nW for (**b**). The PL intensity for $\theta = 20°$ heterobilayer is still in the linear regime for powers less than 100 nW. Aside from the excitation power, the data from all three twist angles were acquired in identical experimental conditions to facilitate their comparison.

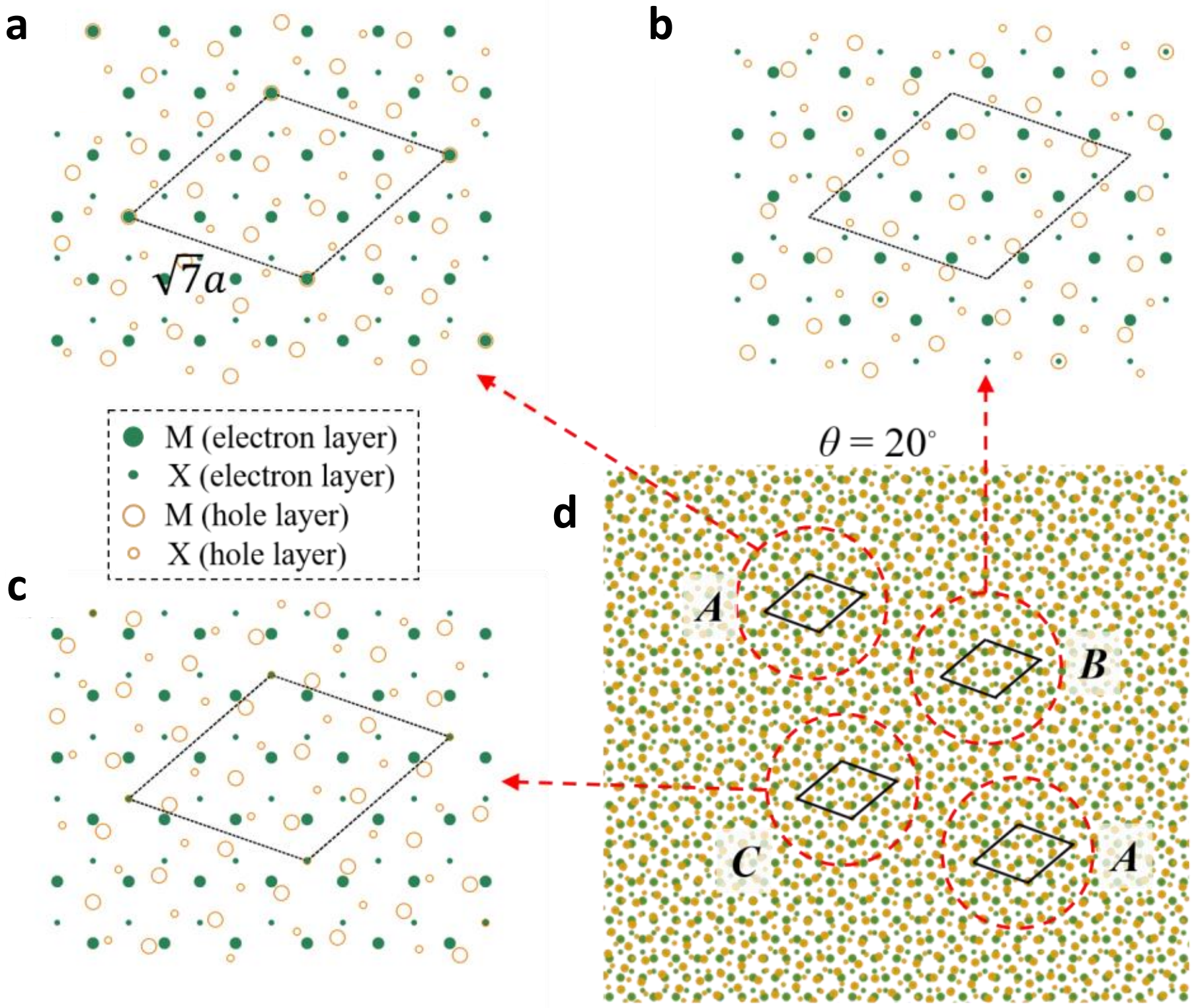

**Extended Data Fig. 8 | Heterobilayer moiré pattern for twist angle close to 21.8°. a-c**, Three high-symmetry stacking patterns under the commensurate 21.8° twist angles for two layers with the same lattice constant $a$. The dashed diamonds give the smallest supercells, whose periodicity is $\sqrt{7}a$. The large (small) solid green dots denote the metal or M (chalcogen or X) sites in the electron layer, while the large (small) empty orange dots denote the metal or M (chalcogen or X) sites in the hole layer. (a) is the stacking where two M sites in different layers horizontally overlap. (b) is the stacking where two hexagon centers (h sites) in different layers horizontally overlap. (c) is the stacking where two X sites in different layers horizontally overlap. Because M, X and h are the $C_3$ rotation centers of the monolayers, the stacking in (a), (b) and (c) are all $C_3$ symmetric. **d**, Illustration of the moiré pattern when the twist angle slightly deviates from 21.8° ($\theta$ = 20° here). Red circles marked by ***A***, ***B*** and ***C*** are the local regions nearly indistinguishable from the 21.8° commensurate pattern in (a), (b) and (c), respectively. The black diamonds are the $\sqrt{7}a \times \sqrt{7}a$ supercells of the local regions.

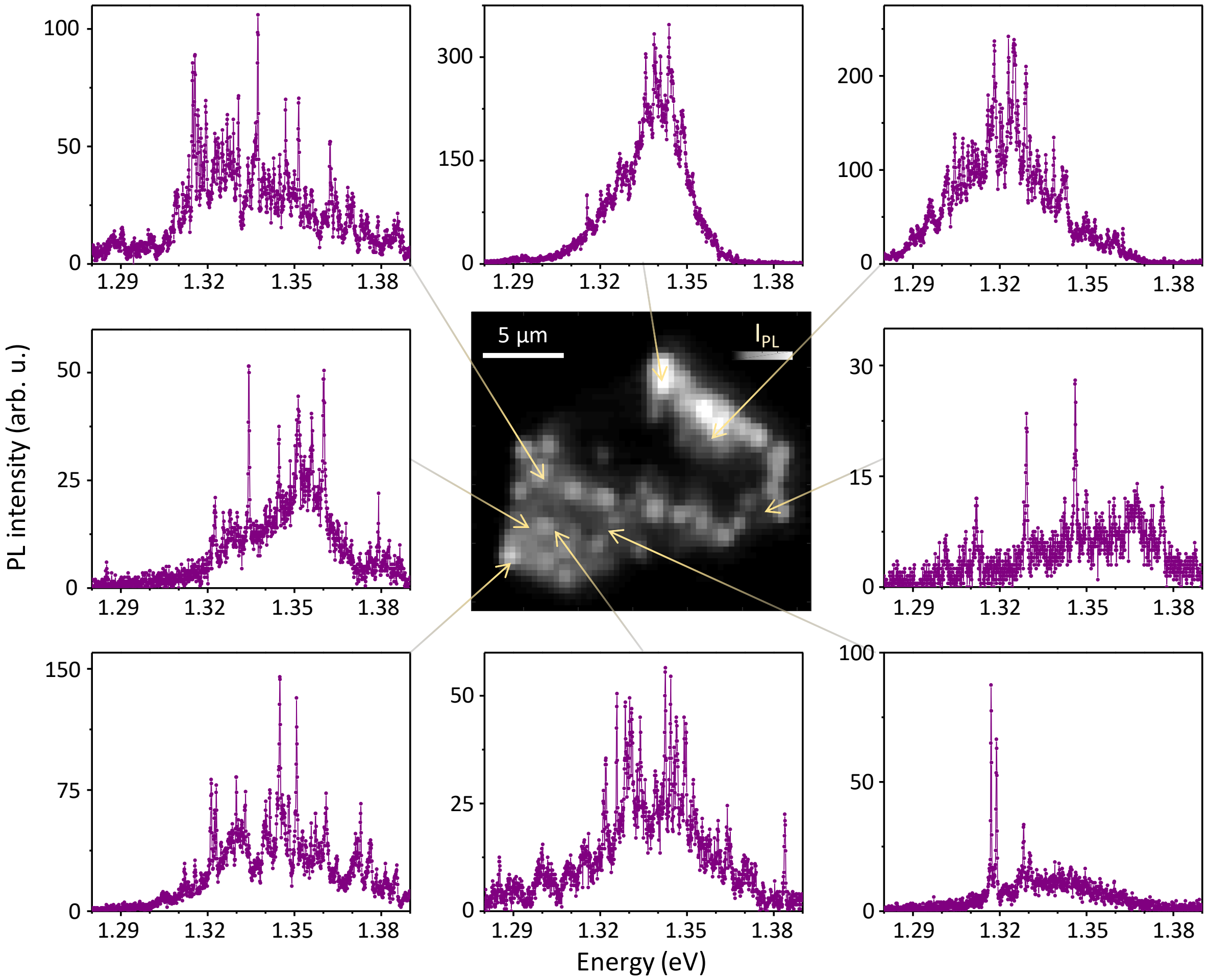

**Extended Data Fig. 9 | Spatial homogeneity of interlayer exciton PL.** The center image displays a spatial map of the integrated PL for the $\theta = 2°$ heterobilayer and the plots along the outside show selected interlayer exciton PL spectra at different sample locations. The PL is bright across a wide region of the sample, except along one edge of a crack that runs through the sample (note correspondence to sample height map in Fig. 1c of main text). Narrow-line PL emission is a general feature of the PL spectra across the sample, as seen in the selected spectra from spatially distinct regions of the sample. The number of narrow-line peaks, their intensities, and their energy distribution is inhomogeneous across the sample. The origin of this inhomogeneity is an important topic for further study.

# Supporting Discussion for

## Signatures of moiré-trapped valley excitons in $MoSe_2/WSe_2$ heterobilayers

**Authors:** Kyle L. Seyler[*1], Pasqual Rivera[*1], Hongyi Yu[2], Nathan P. Wilson[1], Essance L. Ray[1], David G. Mandrus[3,4,5], Jiaqiang Yan[3,4], Wang Yao[#2], Xiaodong Xu[#1, 6]

### S1: Interlayer exciton $g$-factors

As discussed in the main text (Figs. 4a, b), the Zeeman shift of the conduction or valence band edges has three contributions: spin ($\Delta_s = 2s_z\mu_B B$), atomic orbital ($\Delta_a = l_i\mu_B B$), and a valley contribution ($\Delta_v = \tau\alpha_i\mu_B B$) from the Berry phase effect. Here, $\tau = \pm 1$ is valley index, $s_z = \pm 1/2$ is electron spin index, $\mu_B$ is Bohr magneton. $\alpha_c = m_0/m_e^*$ and $\alpha_v = m_0/m_h^*$ are the valley $g$-factors for the conduction and valence bands, respectively, according to a simplified 2-band **k.p** description of the band edges[1] with $m_e^*$ ($m_h^*$) the electron (hole) effective mass and $m_0$ the free electron mass. The two spin-split conduction bands have close effective masses, so we expect their $\alpha_c$ values to be about the same. $l_c = 0$ and $l_v = 2\tau$ are the magnetic quantum number for the atomic orbitals at the conduction and valence band edges.

For spin-conserved optical transitions, the electron and hole spin contributions to the interlayer exciton Zeeman shift cancel with each other, and only $\Delta_a$ and $\Delta_v$ are important. For near 0° stacking, the bright interlayer exciton has two configurations with valley pairing $(\tau_c, \tau_v) = (+,+)$ and $(-,-)$, whose Zeeman shifts are $-(2+\alpha_v-\alpha_c)\mu_B B$ and $(2+\alpha_v-\alpha_c)\mu_B B$, respectively. For near 60° or 21.8° stacking, the bright interlayer exciton has two configurations with the valley pairing $(\tau_c, \tau_v) = (-,+)$ and $(+,-)$, whose Zeeman shifts are $-(2+\alpha_v+\alpha_c)\mu_B B$ and $(2+\alpha_v+\alpha_c)\mu_B B$, respectively. From the measured exciton g-factors, we have

$$2|2+\alpha_v-\alpha_c| \approx 6.7,$$
$$2|2+\alpha_v+\alpha_c| \approx 15.9,$$

which lead to $\alpha_v \approx 3.65$ in $WSe_2$ and $\alpha_c \approx 2.3$ in $MoSe_2$.

The sign of the $g$-factor depends on how the Zeeman splitting is defined. If the Zeeman splitting is defined as the energy of the $(\tau_c, +)$ interlayer exciton minus that of $(-\tau_c, -)$, i.e. according to the hole valley index, then the obtained g-factors should always be negative, i.e., given by $-2(2+\alpha_v-\alpha_c)$ or $-2(2+\alpha_v+\alpha_c)$. However, in the experiments, our observable to distinguish the time-reversal pair of valley configurations is the PL polarization only, so the Zeeman splitting is defined here as the $\sigma^+$ PL peak energy minus $\sigma^-$ PL peak energy. The valley optical selection rule, namely whether $(\tau_c, +)$ emits $\sigma^-$ or $\sigma^+$ photon, then determines the sign of the $g$-factor.

As pointed out in earlier works[2–5], the circularly polarized valley optical selection rules for interlayer excitons depend on the interlayer registry and hence the location in a moiré supercell. Taking the $(+,+)$ interlayer exciton in a near 0° moiré pattern as an example, it emits a $\sigma^+$ ($\sigma^-$) photon at the $\boldsymbol{A}$ ($\boldsymbol{B}$) trapping sites of the moiré potential which has the interlayer registry $R_h^h$ ($R_h^X$). Here, $R_h^\mu$ denotes a 0° lattice-matched stacking, with the $\mu$ site of the electron layer vertically aligned with the hexagon center ($h$) of the hole layer. Interlayer excitons trapped at $\boldsymbol{A}$ and $\boldsymbol{B}$ sites then have $g$-factors with minus and plus signs, respectively.

For the sample with 57° (2°) twist angle, our measured $g$-factor of -15.9 (+6.7) implies that the PL emission is from interlayer excitons localized at $\boldsymbol{A}$ ($\boldsymbol{B}$) trapping site of the moiré supercell. The different signs of the $g$-factors are also consistent with the co-circular PL polarization for 57° and 20° samples and cross-circular polarization for the 2° sample. For example, when exciting at the monolayer exciton resonance in $WSe_2$ with a $\sigma^+$ laser, most of the excited holes will reside in the K valley. The majority interlayer exciton species is then the $(+,+)$ valley pairing in the 2° sample and $(-,+)$ valley pairing in 57° and 20° samples. At low temperature, the valley-polarized interlayer excitons will relax to the local energy minima, that is, $\boldsymbol{A}$ trapping sites in 57° and 20° samples ($\boldsymbol{B}$ sites in 2° sample), which emit $\sigma^+$ ($\sigma^-$) circularly polarized PL, i.e. co-polarized (cross-polarized) with the excitation laser, consistent with the experiment.

In the above analysis, we considered only the excitons of the spin-conserved optical transitions, also known as the spin-singlet excitons. In samples with near 0° twist angle, we expect the PL emission is always from the spin-singlet interlayer exciton that is bright and has the lowest energy. However, in samples with near 60° or 21.8° twist angles, the lowest energy interlayer excitons have the spin-triplet configuration due to the $MoSe_2$ conduction band spin alignment. In heterobilayers, despite its spin-flip nature, the optical dipole of the spin-triplet exciton can be comparable to that of the spin-singlet one[5]. It is therefore possible that the PL emission from these samples arise from the spin-flip optical transitions of the spin-triplet exciton. In such case, the spin contribution to the exciton Zeeman shift are finite, and the total shifts become $-(4+\alpha_v+\alpha_c)\mu_B B$ and $(4+\alpha_v+\alpha_c)\mu_B B$ respectively for $(\tau_c,\tau_v)=(-,+)$ and $(+,-)$. The measured $g$-factor then gives us the equation: $2|4+\alpha_v+\alpha_c|\approx 15.9$. Combined with $2|2+\alpha_v-\alpha_c|\approx 6.7$ from the 0° samples, these yield $\alpha_v\approx 2.65$ in $WSe_2$ and $\alpha_c\approx 1.3$ in $MoSe_2$. The spin-triplet interlayer excitons also have circularly polarized valley optical selection rules at the moiré trapping sites as dictated by the rotational symmetry[5], and our previous discussion about $g$-factor signs still applies.

## S2: Heterobilayer moiré pattern for a twist angle close to 21.8°

For two transition metal dichalcogenide layers with the same lattice constant $a$, a twist angle of 21.8° corresponds to a commensurate pattern with the smallest supercell size ($\sqrt{7}a\times\sqrt{7}a$). Like the 0° or 60° lattice-matched case, here the commensurate 21.8° bilayer also has an interlayer translation degree of freedom, which defines the different stacking configurations. In Extended Data Figs. 8a-c, we show three $C_3$-symmetric stacking at 21.8°. In (a), the stacking corresponds to a metal site (M) in the electron layer overlapped with a metal site in the hole layer. In (b), a hexagon center (h) in the electron layer overlaps with a hexagon center in the hole layer. In (c), a chalcogen site (X) in the electron layer overlaps with a chalcogen site in the hole layer. In the 21.8° heterobilayer of the stacking in (a), (b) and (c), a spin-singlet interlayer exciton with the valley pairing $(\tau_c,\tau_v)=(-,+)$ then has the $C_3$ quantum numbers +1, -1, and 0 respectively[3,5]. This implies that for the stacking in (a), the $(-,+)$ interlayer exciton can emit a $\sigma^+$ circularly polarized photon, whereas for the stacking in (b), it can emit a $\sigma^-$ circularly polarized photon.

The slight deviation $\delta\theta$ of the twist angle from the commensurate 21.8° will give rise to a concatenated moiré pattern. Different local regions correspond to 21.8° commensurate stacking of different interlayer translation, as shown in Extended Data Fig. 8d, where the $\boldsymbol{A}$, $\boldsymbol{B}$ and $\boldsymbol{C}$ locales correspond to the stacking pattern in (a), (b) and (c), respectively. The moiré periodicity is given by $\frac{a}{\sqrt{7}\delta\theta}$. Meanwhile the stacking-dependent interlayer hopping can lead to the periodic change of electronic band gaps[6], which gives rise to a spatially modulated excitonic potential $E(\mathbf{R})$, where $\mathbf{R}$ is the center position of the localized interlayer exciton. At locations $\boldsymbol{A}$, $\boldsymbol{B}$ or $\boldsymbol{C}$, the $C_3$ symmetry requires $\nabla_{\mathbf{R}}E(\mathbf{R}) = 0$, which means they correspond to energy extrema. Thus, an interlayer exciton will be localized at $\boldsymbol{A}$, $\boldsymbol{B}$ or $\boldsymbol{C}$ trapping sites in near 21.8° moiré pattern.